\documentclass[aps,prl,superscriptaddress]{revtex4}
\usepackage[dvips]{graphicx}
\usepackage{indentfirst}
\usepackage{bm}
\usepackage{enumerate}
\def\be{\begin{equation}}
\def\ee{\end{equation}}
\def\bea{\begin{eqnarray}}
\def\eea{\end{eqnarray}}
\newcommand{\ket}[1]{\mbox{$|#1\rangle$}}
\newcommand{\bra}[1]{\mbox{$\langle#1|$}}
\newcommand{\avg}[1]{\mbox{$\langle#1\rangle$}}

\newcommand{\op}[2]{\mbox{$\hat{#1}_{#2}$}}

\def\Aeff{A_{\scriptsize\textrm{eff}}}
\begin{document}
\title{Crystallization of strongly interacting photons in a nonlinear optical fiber}

\author{D.E. Chang}
\affiliation{Department of Physics, Harvard University, Cambridge,
Massachusetts 02138}
\author{V. Gritsev}
\affiliation{Department of Physics, Harvard University, Cambridge,
Massachusetts 02138}
\author{G. Morigi}
\affiliation{Grup d'Optica, Department de Fisica, Universitat
Autonoma de Barcelona, 08193 Bellaterra, Spain}
\author{V. Vuleti\'{c}}
\affiliation{Department of Physics, MIT-Harvard Center for
Ultracold Atoms, and Research Laboratory of Electronics,
Massachusetts Institute of Technology, Cambridge, Massachusetts
02139}
\author{M.D. Lukin}
\affiliation{Department of Physics, Harvard University, Cambridge,
Massachusetts 02138}
\author{E.A. Demler}
\email{demler@physics.harvard.edu} \affiliation{Department of
Physics, Harvard University, Cambridge, Massachusetts 02138}

\date{\today}

\begin{abstract}
Understanding strongly correlated quantum systems is a central
problem in many areas of physics. The collective behavior of
interacting particles gives rise to diverse fundamental phenomena
such as confinement in quantum chromodynamics, phase transitions,
and electron fractionalization in the quantum Hall regime. While
such systems typically involve massive particles, optical photons
can also interact with each other in a nonlinear medium. In
practice, however, such interactions are often very weak. Here we
describe a novel technique that allows the creation of a strongly
correlated quantum gas of photons using one-dimensional optical
systems with tight field confinement and coherent photon trapping
techniques. The confinement enables the generation of large,
tunable optical nonlinearities via the interaction of photons with
a nearby cold atomic gas. In its extreme, we show that a quantum
light field can undergo fermionization in such one-dimensional
media, which can be probed via standard photon correlation
measurements.
\end{abstract}
\maketitle

The idea of using nonlinear effects to create optical systems with
unusual properties has been pursued for several decades, and
fascinating advances have made in recent
years~\cite{boyd92,haroche01,miller05,walther06,fleischhauer05}.
The focus of such work has mostly been on systems containing a
small number of photons. At the same time, the effects of strong
correlations manifest themselves most dramatically in many-body
systems, often resulting in new states of matter with properties
that are very different from those of the underlying particles.
One famous example is the Tonks-Girardeau~(TG) regime of
interacting bosons in one dimension, in which strong interactions
between particles lead to an effective ``fermionization" of
bosons~\cite{TG,TG2}. This unusual regime has been explored in
several condensed matter systems and in recent experiments with
ultracold atoms in one dimensional traps~\cite{NBW,NBW3}. In this
article we investigate the feasibility of creating and detecting a
TG gas of photons. This system would correspond to nonlinear
quantum optics in its extreme, in which individual photons behave
as impenetrable particles. In this limit, an optical pulse
separates into non-overlapping wavepackets of individual photons,
and a ``crystal of photons" can be created.

Several papers have recently considered the possibility of phase
transitions involving photons, focusing on large systems of
coupled optical cavities~\cite{others,others2,others3,others4}.
The present study extends this important work along two principal
directions.  First, while dramatic progress has been achieved in
controlling individual atoms and photons in single
cavities~\cite{haroche01,miller05,walther06}, the complex
architecture of coupled cavities proposed
in~\cite{others,others2,others3,others4} represents a considerable
experimental challenge. In contrast, the approach described here
to realize a TG gas of photons involves currently available
experimental techniques. Second, we show how signatures of a
strongly interacting photon gas can be detected using standard
quantum optical measurements. This potentially enables many new
applications and allows one to study even more exotic phenomena
involving interacting particles using readily controlled photons
with tunable interactions.

\section{Strongly interacting photons in one dimension: the system}

Recently, much effort has been directed toward realizing
single-mode optical waveguides where the guided photons can be
tightly confined in the transverse directions to an area $\Aeff$
near or below the diffraction limit. Such confinement is desirable
in part because it allows for a large interaction strength between
single photons and nearby coupled atoms. Specific systems that
have recently been explored in this context include tapered
optical fibers~\cite{nayak07}, hollow-core photonic crystal
fibers~\cite{knight03,ghosh05}, and surface plasmons on conducting
nanowires~\cite{chang06,akimov07}.

\begin{figure*}[p]
\begin{center}
\includegraphics[width=18cm]{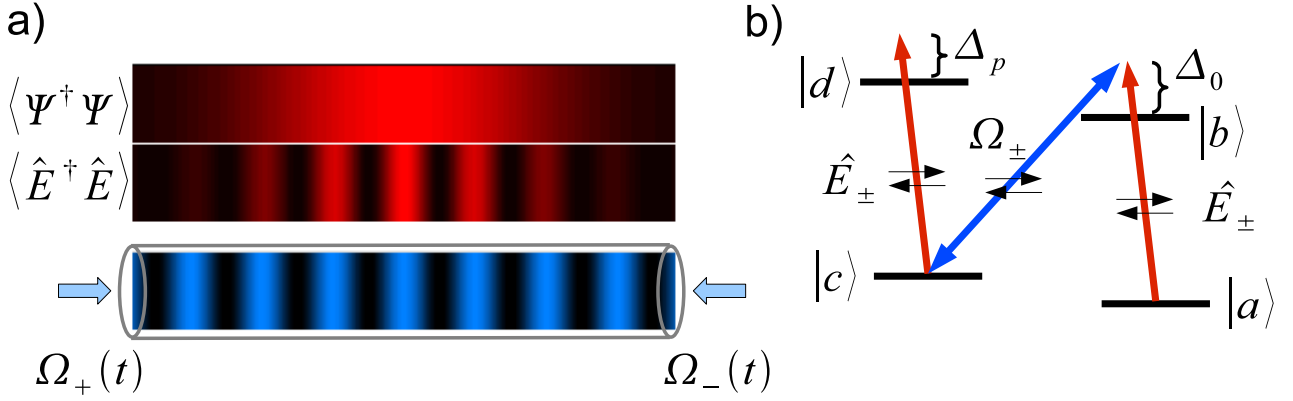}
\end{center}
\caption{\textbf{Illustration of fields and atoms comprising the
system} a) Schematic of fields inside the waveguide, whose axis of
propagation corresponds to the horizontal axis. The control beams
$\Omega_{\pm}(t)$~(shown in blue) create a standing wave inside
the waveguide, which forms a Bragg grating that traps a quantum
optical field inside the medium~(intensity
$\langle\hat{E}^{\dagger}\hat{E}\rangle$ shown in red).  The
optical field couples to spin-wave excitations in the medium,
resulting in collective polariton excitations whose density
$\avg{\Psi^{\dagger}\Psi}$ is also plotted. b) Schematic of the
four-level atomic configuration and coupling between levels and
fields used in our system.\label{fig:fourlevels}}
\end{figure*}

The propagation and interaction of photons in such a medium can be
controlled by interfacing them with atoms using quantum optical
techniques such as Electromagnetically Induced
Transparency~(EIT)~\cite{fleischhauer05}. In particular, our
scheme relies on resonantly enhanced optical nonlinearities with
low losses~\cite{schmidt96} using EIT and the creation and
trapping of stationary pulses of light in the medium using
spatially modulated control fields~\cite{bajcsy03}~(see
Fig.~\ref{fig:fourlevels}a). As will be shown, the dynamical
evolution of the photonic system is governed by an equation that
has the form of the Nonlinear Schr\"{o}dinger Equation~(NLSE),
where the signs and strengths of the effective mass and
interaction can be controlled using external fields. Under
conditions where the nonlinear interaction is large and
effectively \textit{repulsive}, the TG regime of photons can be
achieved.

Specifically, we consider the propagation of counter-propagating
quantum fields, characterized by operators $\hat{E}_{\pm}$, inside
a single-transverse mode optical waveguide and interacting with
cold atoms in the four-level configuration shown in
Fig.~\ref{fig:fourlevels}b. The fields $\op{E}{\pm}$ couple the
ground state $\ket{a}$~(in which the system is initialized) to
excited state $\ket{b}$ with a strength given by $g$, while
metastable state $\ket{c}$ and $\ket{b}$ are coupled by classical,
counter-propagating control fields $\Omega_{\pm}(t)$~(we assume
these fields are also guided). Here, the lambda configuration
consisting of states $\ket{a,b,c}$ comprise the typical EIT setup.
In particular, in the case where the quantum and classical fields
propagate only in one direction, the quantum field can be
dynamically and reversibly mapped into a stationary spin-wave
excitation by turning the control field $\Omega(t)$ to zero
adiabatically. On the other hand, as shown in~\cite{bajcsy03}, by
creating a standing wave formation with the control
fields~($\Omega_{+}(t)=\Omega_{-}(t)$), the quantum fields can be
effectively stopped while simultaneously maintaining a non-zero
photonic component of the excitation~(see
Fig.~\ref{fig:fourlevels}a). Intuitively, the standing wave
pattern forms a set of reflection gratings that trap the photonic
excitations through multiple scattering. The nonlinear response is
introduced via an additional state $\ket{d}$ that is coupled to
$\ket{c}$ by the quantum fields~(for simplicity we assume that
this coupling strength is also given by $g$). When these fields
are far off resonance, this coupling results in an ac Stark shift
of state $\ket{c}$ whose magnitude is proportional to the
intensity of the quantum field~\cite{schmidt96}. This gives rise
to an intensity-dependent refractive index, or nonlinear
susceptibility, that in turn influences the evolution of the
quantum fields. While similar schemes for realizing nonlinear
optics in atomic media have been previously
explored~\cite{bajcsy05}, their implementation in waveguides with
tight confinement is unique because they can constitute a true
one-dimensional system, and because the strong transverse
localization enables large nonlinear interactions.

\section{The Lieb-Liniger model with stationary pulses of light}

Considerable theoretical literature exists discussing the
connection between Maxwell's equations for a nonlinear medium in
one dimension and the Lieb-Liniger Model (LLM) describing
interacting massive particles~\cite{NLSE2,Drummond}. However, most
of this work has focused on the regime of attractive interactions.
Interest in this regime stems from the observation that
interaction with basic two-level atomic systems leads to photon
bunching, since it is easier for many photons to pass through an
atom without being absorbed~\cite{chang07}. The main feature of
the attractive regime is the formation of solitons~\cite{NLSE2}.
This regime of the LLM has few quantum mechanical features and can
be described by classical equations of motion. Indeed, typical
optical solitons contain on the order of a million photons, and
therefore a quasi-classical description is very
appropriate~\cite{MK}. On the other hand, the {\it repulsive} case
is intrinsically ``quantum mechanical", as one needs to keep track
of correlations at the level of individual particles. Solutions
based on perturbation theory or classical equations of motion
break down and one needs to use non-perturbative approaches such
as the Bethe ansatz solution~\cite{LL,BIK,CC,CC2} or the Luttinger
liquid formalism~\cite{Haldane}. Thus, formation of a TG gas of
photons is fundamentally a collective {\it many-body} effect.

We now derive an evolution equation for the fields of the system
illustrated in Figure~\ref{fig:fourlevels}. Following the methods
of~\cite{fleischhauer00,bajcsy03,bajcsy05}, we define dark-state
polariton operators $\Psi_{\pm}$ describing the coupled photonic
and spin-wave excitations, which in the slow-light limit are given
approximately by
$\Psi_{\pm}=g\sqrt{2\pi{n_z}}\op{E}{\pm}/\Omega_{\pm}$, where
$n_z$ is the density of atoms coupled to the waveguide~(the
density is assumed to be uniform). We specialize to the case when
$\Omega_{\pm}(t)=\Omega(t)$. We further assume that the quantum
and control fields vary slowly in time, such that the fast-varying
atomic operators can be adiabatically eliminated, while the
remaining slowly-varying operators are solved in the adiabatic
limit. Subsequently inserting these solutions into the
Maxwell-Bloch equations describing evolution of the quantum fields
yields an effective NLSE for the photon gas~(see Methods),
\be
i\partial_{t}\Psi(z,t)=-\frac{1}{2m_{\scriptsize\textrm{eff}}}\partial_{z}^{2}\Psi(z,t)+2\tilde{g}\Psi^{\dagger}(z,t)\Psi^{2}(z,t),\label{eq:NLSE}
\ee
where
\bea\label{LLconstants}
\Psi=\frac{\left(\Psi_{+}+\Psi_{-}\right)}{2},\qquad
m_{\scriptsize\textrm{eff}}=-\frac{\Gamma_{1D}
n_{z}}{4\Delta_{0}v_{g}},\qquad 2\tilde{g}=\frac{\Gamma_{1D}
v_{g}}{\Delta_{p}}. \eea
Here $\Gamma_{1D}=4\pi{g^2}/v$ is the spontaneous emission rate of
a single atom into the waveguide modes, where $v$ is the velocity
of these modes at the atomic resonance frequency in an empty
waveguide, while $v_g{\approx}v\Omega^{2}/({\pi}g^{2}n_z)$ is the
group velocity of untrapped pulses under EIT conditions.
$\Delta_{0}$ and $\Delta_p$ are the one-photon detunings of the
fields $\hat{E}_{\pm}$ from the transitions $\ket{a}$-$\ket{b}$
and $\ket{c}$-$\ket{d}$, respectively~(see
Fig.~\ref{fig:fourlevels}b), and $g{\sim}1/\sqrt{\Aeff}$ is the
single-photon, single-dipole interaction matrix element. In
principle the full dynamics of the field will also include losses
and higher-order terms, and the conditions under which such terms
can be neglected are described in Methods.

Eq.~(\ref{eq:NLSE}) determines the evolution of a quantum field
$\Psi(z,t)$ as derived from the Hamiltonian of the Lieb-Liniger
model,
\bea\label{LL} H=\hbar\int
dz[\frac{1}{2m_{\scriptsize\textrm{eff}}}\partial_{z}\Psi^{\dag}(z)\partial_{z}\Psi(z)+
\tilde{g}\Psi^{\dag}(z)\Psi^{\dag}(z)\Psi(z)\Psi(z)]. \eea
The first term on the right describes the kinetic energy in one
dimension of bosons with mass $m_{\scriptsize\textrm{eff}}$, while
the second term describes a contact interaction potential. The
quantum field $\Psi(z,t)$ satisfies the usual equal-time bosonic
commutation relations,
$[\Psi(z,t),\Psi^{\dag}(z',t)]=\delta(z-z')$. The behavior and
properties of this system can be effectively characterized by a
single dimensionless parameter
\bea \gamma = \frac{m_{\scriptsize\textrm{eff}}\tilde{g}}{n_{ph}}
=-\frac{\Gamma_{1D}^{2}}{8\Delta_{0}\Delta_{p}}\frac{n_{z}}{n_{ph}},\label{eq:gamma}
\eea
which physically corresponds to the ratio of the interaction and
kinetic energies.  Here $n_{ph}$ is the density of photons at the
center of the pulse. When $\gamma<0$, the interaction with the
atoms induces an effective attraction between photons. As
discussed above, this is responsible, \textit{e.g.}, for the
formation of bound states~(solitons) when a large number of
photons are present. On the other hand, for $\gamma>0$ the regime
of effective {\it repulsion} between photons is realized. The
special limit $\gamma\rightarrow\infty$ is called the
Tonks-Girardeau~(TG) regime~\cite{TG,TG2}. From
Eq.~(\ref{LLconstants}), one finds that the regime of repulsion
can be achieved if exactly {\it one} of the detuning parameters
$\Delta_{0}$ or $\Delta_{p}$ is negative. It is also clear that
only the attractive regime can be reached using a two-level
system, since $\Delta_{0}\equiv\Delta_{p}$ in this situation.
Finally, we note that
$\gamma{\propto}\Gamma_{1D}^2\propto\Aeff^{-1}$, which underscores
the importance of tight mode confinement in reaching the strongly
interacting regime.

One useful feature of this realization of the NLSE is that the
parameters $m_{\scriptsize\textrm{eff}}$ and $\tilde{g}$ can be
dynamically tuned by varying different parameters of the system.
Specifically, both are functions of detuning, and thus $\gamma$
can be altered by changing either the control field frequencies or
by externally manipulating the energies of the atomic levels.  As
described below, this tunability facilitates the creation of novel
photonic states.

\section{Preparation and detection of strongly correlated photon gas}
\label{preparation}

The process by which novel photonic states can be prepared and
detected consists of three basic steps -- loading of the pulse,
controlled evolution under the NLSE, and readout of the final
photonic state -- each of which we now describe. During the
initial loading process, a resonant optical pulse, given by,
\textit{e.g.}, a coherent state, is incident from one direction.
It is injected into the waveguide at the same time that the
co-propagating control field~(say $\Omega_{+}(t)$) is turned on.
During the loading procedure the counter-propagating control field
$\Omega_{-}(t)$ is off. This describes the usual situation in EIT,
where the input field becomes spatially compressed upon entering
the medium and propagates with a variable group velocity
$v_{g}{\sim}v\Omega_{+}^2(t)/(g^2{n_z})$.  Once the pulse
completely enters the medium, $\Omega_{+}(t)$ is adiabatically
turned to zero, reversibly converting the photonic excitation into
a pure spin-wave excitation in the atomic
medium~\cite{fleischhauer00}.  Under certain conditions this input
pulse can be stored with minimal distortion relative to the
initial field~\cite{fleischhauer00}, such that all relevant
properties~(\textit{e.g.}, field correlations) remain constant
during the loading.  Following the initial storage, both control
fields are then adiabatically switched on, with
$\Omega_{\pm}=\Omega(t)$, during which the pulse becomes trapped
and evolves under Eq.~(\ref{eq:NLSE}).  The parameters
$m_{\scriptsize\textrm{eff}}$ and $\tilde{g}$ can be changed in
time during the evolution to reach the final state of interest.
Once this state has been achieved, the pulse is released by
turning one of the control fields~(say $\Omega_{-}$) off, thereby
allowing the pulse to propagate undistorted until it exits the
waveguide~\cite{bajcsy03}. During this readout of the pulse, any
spatial correlations that formed while evolving under the NLSE are
directly mapped into temporal correlations~(at a common point in
space) in the outgoing field, which can be measured using standard
quantum optical techniques.

A characteristic signature of a strongly interacting gas is the
appearance of Friedel oscillations~\cite{friedel} in the
normalized second-order correlation function, $g^{(2)}(z,z')=
\langle I(z) I(z') \rangle/(\avg{I(z)}\avg{I(z')})$, where
$I(z)=\Psi^{\dagger}(z)\Psi(z)$ is the stationary pulse intensity
prior to release. In particular, if the photonic state is close to
the ground state of the LLM, $g^{(2)}(z,z')$ contains an
oscillating part that behaves as $\sim\cos (2k_{F} (z-z'))$~(see
Methods), where $k_F=\pi{n_{ph}}$ is the Fermi momentum in the TG
limit. In the TG limit, the ground-state correlation function
takes on a simple form given by Lenard's formula~\cite{Lenard2},
\bea g^{(2)}_{TG}(z,z')=1-\left(\frac{\sin k_{F}(z-z')}{k_{F}
(z-z')}\right)^{2}. \eea
The $2k_{F}$-oscillations are a direct manifestation of
``fermionization'' of bosons.  We note that Friedel oscillations
are more than simple anti-bunching in that they indicate real
crystal correlations.  In particular, one cannot predict the
position of an individual photon, but knowing the position of one
photon, other photons are likely to follow at well-defined
distances determined by the average photon density.  These
correlations are predicted to decay relatively slowly in space.

The usual definition of the TG regime of the LL model is given
with respect to the equilibrium state. However, in our system the
evolution of the initial coherent state under the NLSE inherently
involves non-equilibrium quantum dynamics, and hence one must
specify the conditions under which one can obtain a strongly
interacting state of photons that is close to the ground state of
the LLM and exhibits its characteristic features, such as Friedel
oscillations. Specifically, we consider the evolution of a pulse
under Eq.~(\ref{eq:NLSE}), where the atomic parameters are varied
such that the effective mass $m_{\scriptsize\textrm{eff}}$ is
constant in time, while the interaction strength $\tilde{g}(t)$
increases exponentially.  From Eq.~(\ref{eq:gamma}), the resulting
time-dependent interaction parameter can be written in the form
$\gamma=\gamma_{0}e^{\beta\omega_{F}t}$, where $\beta$ is a
dimensionless parameter characterizing the rate of increase, and
$\omega_{F}{\sim}n_{ph}^2/m_{\scriptsize\textrm{eff}}$ corresponds
to the ``Fermi energy'' at the center of the pulse before
expansion.  We assume that $\gamma_{0}{\ll}1$, so that the system
initially consists of non-interacting photons, and interactions
are gradually switched on to reach the regime with $\gamma \gg 1$.
Here we will not discuss the effect of a time-dependent mass,
although it can be included using a similar analysis. We also note
that the related problem involving the sudden switching of the
interaction strength from zero to a large value has been studied
in~\cite{RDG}.

The non-equilibrium dynamics of the LLM model has previously been
studied in the context of ultracold atoms and has focused either
on changing the interaction strength in a system with uniform,
constant density~\cite{caz,PG}, or on the expansion of particles
in a system with constant interaction~\cite{muramatsu,MG}. In our
system both processes take place, but in the experimentally
relevant regime of a large photon number there is a separation of
time scales that simplifies the analysis.  We consider an initial
pulse containing $N_{ph}{\sim}n_{ph}z_{0}$ photons with spatial
extent $z_0$ at $t=0$. Our discussion applies for a general pulse
shape and thus we need not specify it.

There are three distinct regimes in the time evolution of the
pulse: i) Interactions are weak and the photons expand freely due
to dispersion; ii) Interactions begin to dominate over the kinetic
energy. However, the system is still in the weakly interacting
regime with $\gamma<1$, in which case the expansion is
hydrodynamic~\cite{CK,CK2}; iii) The system reaches the strongly
interacting regime with $\gamma>1$, and the expansion resembles
that of fermionized bosons~\cite{muramatsu,MG}. A simple
analysis~(see Methods) shows that by the time the system reaches
regime iii), the relative change in the photon density at the
center of the pulse is only of the order of $\frac{1}{\beta^{2}
N_{ph}^2}$. Hence for $\beta N_{ph}
>>1$ one can assume that the turning on of interactions takes place
at a constant density and any subsequent expansion takes place in
the TG regime.

We now consider the effect of non-adiabaticity on correlation
functions such as $g^{(2)}(z,z')$. By analogy with ultracold
atoms, we introduce a chemical potential, which in regimes i) and
ii) is given by $\mu(t) \approx \tilde{g}(t) n_{ph}$. We can
approximately separate the turning on of interactions into two
stages, which both take place during the regimes i) or ii). In the
first stage, $\dot{\mu} > \mu^2$ and the evolution is diabatic. In
the second stage, $\dot{\mu} < \mu^2$ and the evolution is
essentially adiabatic.  At the time $t_{\it adiab}$ separating the
two regimes, the interaction parameter is given by $\gamma(t_{\it
adiab}){\sim}\beta$. The first stage can be thought of as an
instantaneous projection of the wavefunction, which gives rise to
a finite density of excitations characterized by an effective
healing length
$\xi_{neq}{\sim}(\mu(t)m_{\scriptsize\textrm{eff}})^{-1/2}$ at
time $t_{\it adiab}$. During the second stage, the number of
excitations does not change. Hence, we find that the finite rate
of change in the interaction strength leads to a finite
correlation length in our system, $\xi_{\it neq}\sim {\beta}
^{-1/2}n_{ph}^{-1}$. For length scales shorter than $\xi_{\it
neq}$, all correlation functions are essentially the same as in
the ground state, whereas for length scales longer than $\xi_{\it
neq}$, correlation functions rapidly decay.  To observe Friedel
oscillations over length scales on the order of the inter-photon
distance, for example, requires that $\beta \lesssim 1$. The
argument presented above can be turned into a quantitative
calculation for the correlation functions following the time
dependent change in the interaction strength~\cite{PG}. This
analysis uses bosonization to treat the LLM and the conclusions
agree with the qualitative picture presented here.

At the end of regime ii) we have a system of ``fermionized'' hard
core photons that should exhibit Friedel oscillations. In regime
iii), the pulse of ``hard-core photons'' expands, but such
spreading does not lead to the suppression of the Friedel
oscillations~\cite{muramatsu,MG}. The problem of expansion of
hard-core photons starting from a general pulse shape has to be
analyzed numerically. However, an explicit analytic solution is
available for the case of a parabolic pulse shape.  Under
realistic conditions this solution, which is discussed in detail
in the Methods section, yields the correlation function shown
in~Fig.~\ref{fig:TG}.

\begin{figure*}[p]
\begin{center}
\includegraphics[width=10cm]{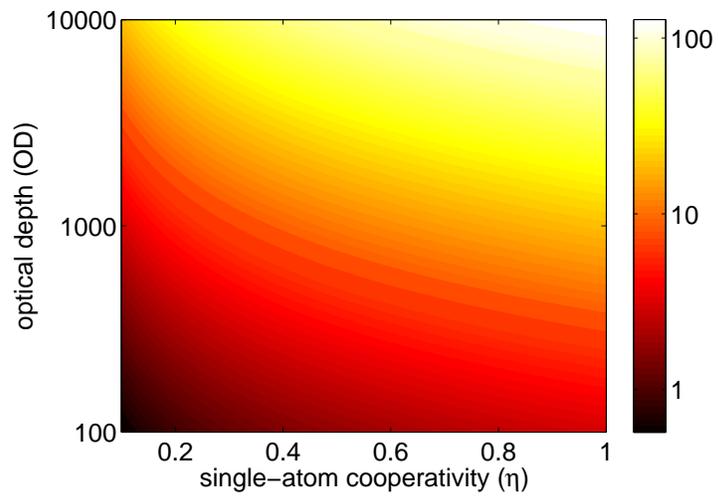}
\end{center}
\caption{\textbf{Maximum interaction parameter} $\gamma_{max}$ as
functions of optical depth and single-atom cooperativity,
optimized over the detuning $\Delta_{0}$.  $\gamma_{max}$ is
plotted for fixed values of $\beta=1$, $\gamma_{0}=0.1$, and
$N_{ph}=10$. It is evident that the TG regime for photons can be
approached by increasing either the cooperativity or optical
depth.\label{fig:gammamax}}
\end{figure*}

In principle, during the evolution one must also consider the
effects of photon losses, which set a maximum evolution time
$t_{max}$ and interaction parameter $\gamma_{max}$ that can be
achieved before a substantial fraction of the initial pulse is
dissipated.  As derived in the Supplementary Information~(SI),
\be
\gamma_{max}{\sim}\min\left(\gamma_{0}\exp\left(\frac{\beta|\Delta_0|}{\Gamma}\right),\eta\beta\frac{\Gamma}{|\Delta_0|}\frac{OD}{N_{ph}}\right),
\ee
where $\Gamma$ is the total spontaneous emission rate of states
$\ket{b}$ and $\ket{d}$~(assumed to be equal for simplicity),
which includes the emission rate into the waveguide
modes~($\Gamma_{1D}$) as well as emission into non-guided
modes~(\textit{e.g.}, into free space). We have also defined a
single-atom cooperativity
$\eta=\Gamma_{1D}/\Gamma$~($\eta{\leq}1$) which describes the rate
of emission into the waveguide compared to the total emission, and
defined the optical depth of the medium, $OD={\eta}z_{0}n_{z}$. We
note that $\gamma_{max}$ can be improved by increasing the optical
depth or cooperativity of the system, and optimized by adjusting
the detuning $|\Delta_0|$. The optimal values of $\gamma_{max}$,
as functions of optical depth and cooperativity, are plotted in
Fig.~\ref{fig:gammamax}, for parameters $\beta=1, \gamma_{0}=0.1$,
and $N_{ph}=10$.  One sees that with realistic values of
$OD{\sim}2000$ and $\eta{\sim}0.2$, for example, an interaction
parameter of $\gamma_{max}{\sim}10$ is possible. While photon
losses limit the maximum evolution time, somewhat surprisingly,
the nonlinear losses may also help in bringing the system closer
to the ground state of the LLM model. Specifically, these losses
predominantly remove states in which two photons are close to each
other, which correspond to high energy states of the LLM. Deep in
the TG regime we can estimate the time it takes for the system to
lose the high energy states, $ t_{\it en}^{-1} \approx
\frac{\Gamma}{\Delta_p}\, \tilde{g} n_{ph} $. On the other hand,
the photon loss rate is $ t_{\it los}^{-1} \approx
\frac{\Gamma}{\Delta_p} \,\omega_F$. Thus $t_{\it en}/t_{\it los}
\approx \gamma^{-1} $ and for large $\gamma$ there is a sufficient
time window for the high-energy states to decay before too many
photons are lost.

\section{Outlook}

The above analysis indicates that strongly correlated states of
photons can be controllably prepared and observed in
one-dimensional waveguides.  These techniques are made possible
through strong coupling between the photons and nearby atoms and
the use of quantum optical techniques such as EIT, which allow the
system to be widely tunable. Such photonic states should find
numerous applications in various areas of physics. Crystal
correlations that arise in the fermionized state make it a
promising candidate for applications in metrology and quantum
information. In particular, TG states feature strongly suppressed
photon number fluctuations within a given detection interval. Such
states therefore could be used as an input for sub-shot noise
interferometers~\cite{dowling,kasevich}, or in extension to
schemes for quantum computing~\cite{kok07} or quantum
cryptography~\cite{gisin02} that rely on single photons. Another
exciting direction is quantum simulation of matter Hamiltonians
using optical systems. In the discussion so far, we have
considered photons with only one polarization. Including photons
of different polarizations should be equivalent to adding a spin
degree of freedom to effective matter Hamiltonians. This opens up
exciting prospects for exploring spin charge separation~(see,
\textit{e.g.},~\cite{scs}) and modelling exotic spin systems. It
is also interesting to note that the level structure of the atoms
comprising the medium can vary considerably. This fact can be used
for the study of strongly-correlated systems with non-Abelian
symmetry, similar to the ones realized in multi-channel Kondo
models and quantum chromodynamics. Possible phases and phase
transitions in these models are difficult, if possible at all, to
realize in matter systems. Furthermore, it would be interesting to
consider a situation where interactions between the cold atoms
create non-trivial correlations~\cite{fleischhauer04} and their
effect on the resulting photonic states. Using light to simulate
matter Hamiltonians will give a new meaning to the old idea of the
particle-wave duality.

\section{Methods}

\subsection{Derivation of NLSE for photons}

The Hamiltonian corresponding to the system described in
Fig.~\ref{fig:fourlevels} is given in a rotating frame by
\bea H=-\hbar
n_{z}\int\,dz[\Delta_{0}\sigma_{bb}+\Delta_p\sigma_{dd}+g\sqrt{2\pi}((\sigma_{ba}+\sigma_{dc})(\op{E}{+}e^{ik_0{z}}
+\op{E}{-}e^{-ik_{0}z})+h.c.)\nonumber\\
+((\Omega_{+}(t)e^{ik_{c}z}+\Omega_{-}(t)e^{-ik_{c}z})\sigma_{bc}+h.c.)],\label{eq:H}
\eea
where $\op{E}{\pm}\equiv \op{E}{\pm}(z,t)$ are slowly-varying
operators describing the quantum fields, and
$\sigma_{ij}\equiv\sigma_{ij}(z,t)$ are collective, continuous
operators describing the average of $\ket{i}\bra{j}$ over atoms in
a small but macroscopic region around $z$. For simplicity, we have
assumed equal transition matrix elements $g_{ba}=g_{dc}=g$ between
the quantum fields $\op{E}{\pm}$ and the transitions
$\ket{a}$-$\ket{b}$ and $\ket{c}$-$\ket{d}$, where
$g_{ij}{\sim}\bra{i}r\ket{j}\sqrt{\omega_{ij}/\hbar\epsilon_{0}\Aeff}$.
Note that we have also included a one-photon detuning $\Delta_{0}$
between the quantum fields and transition
$\ket{a}$-$\ket{b}$~(while maintaining two-photon resonance),
whose purpose is to provide a group velocity dispersion or
``effective mass'' term in the field evolution equations. Defining
slowly varying operators
$\sigma_{ab}=\sigma_{ab,+}e^{ik_{0}z}+\sigma_{ab,-}e^{-ik_{0}z}$
and
$\sigma_{cd}=\sigma_{cd,+}e^{ik_{0}z}+\sigma_{cd,-}e^{-ik_{0}z}$,
the Maxwell-Bloch equations describing evolution of the fields
under $H$ are given by
\be
\left(\frac{1}{v}\frac{\partial}{{\partial}t}\pm\frac{\partial}{{\partial}z}\right)\op{E}{\pm}(z,t)=\frac{\sqrt{2\pi}ign_z}{v}\left(\sigma_{ab,\pm}(z,t)+\sigma_{cd,\pm}(z,t)\right),\label{eq:propeq}
\ee
while the usual Langevin-Bloch equations describing evolution of
$\sigma_{ij}$ can be derived following the methods
of~\cite{fleischhauer00,bajcsy03,bajcsy05}. Following these
references, we define polariton operators ${\Psi}_{\pm}$ to
describe the collective excitations of field and spin-wave
coherence $\sigma_{ac}$ that result from coupling with the control
fields, which in the relevant limit that the excitations are
mostly in spin-wave form are given by
$\Psi_\pm=g\sqrt{2\pi{n_z}}\op{E}{\pm}/\Omega_{\pm}$.  To proceed
further, we adiabatically eliminate the Langevin-Bloch equations
for the fast-decaying atomic operators~(\textit{e.g.},
$\sigma_{ab}$ and $\sigma_{cd}$), and slowly-varying operators are
solved in terms of $\Psi_\pm$, discarding higher time derivatives
in the slowly-varying limit. Plugging these results back into
Eq.~(\ref{eq:propeq}), and specializing to the case where
$\Omega_{\pm}(t)=\Omega$, we obtain evolution equations for the
polaritons alone,
\bea \frac{1}{v}\partial_{t}\Psi+\partial_{z}A & = &
-\frac{1}{v_g}\partial_{t}\Psi-\frac{2{\pi}ig^2}{v(2\Delta_{p}+i\Gamma)}\left(2\Psi^{\dagger}\Psi+A^{\dagger}A\right)\Psi+\textrm{noise},\label{eq:ddtS}
\\ \frac{1}{v}\partial_{t}A+\partial_{z}\Psi & = &
-\frac{4\pi{g^2}{n_z}}{v(\Gamma-2i\Delta_0)}A-\frac{2{\pi}ig^2}{v(2\Delta_p+i\Gamma)}\Psi^{\dagger}\Psi{A}+\textrm{noise}.\label{eq:ddtA}
\eea
Here we have defined the symmetric and anti-symmetric combinations
$\Psi=(\Psi_{+}+\Psi_{-})/2$ and $A=(\Psi_{+}-\Psi_{-})/2$, and a
group velocity $v_{g}{\approx}v\Omega^2/(\pi{g^2}{n_z})$ under
which the pulses would propagate were they not trapped.  The total
spontaneous emission rates $\Gamma$~(which include decay into
channels other than the guided fiber modes) from states
$\ket{b},\ket{d}$ are assumed to be identical for simplicity.  We
note that there are also noise operators, which are associated
with the dissipative terms in the equations above.  Because we are
primarily interested in the regime where losses are not
significant, the specific form of these operators is not important
here. With sufficient optical depth, $A$ can be adiabatically
eliminated,
$A{\approx}(2i\Delta_0-\Gamma)v(\partial_{z}\Psi)/(4{\pi}g^2{n_z})$,
where we have assumed that the nonlinear contribution
${\sim}\Psi^{\dagger}{\Psi}A$ is small. Physically, this result
corresponds to a pulse matching phenomenon~\cite{harris93} between
the quantum and control fields, whereby any imbalance between
$\Psi_{+}$ and $\Psi_{-}$ rapidly goes to zero. Substituting the
expression for $A$ back into Eq.~(\ref{eq:ddtS}), and considering
the relevant case where $v_{g}{\ll}v$ yields the NLSE with complex
effective mass and two-body interaction strength,
\bea\label{NLS} i\partial_{t}\Psi=
\frac{(2\Delta_0+i\Gamma){v_g}v}{4{\pi}g^2{n_z}}\partial_{z}^{2}\Psi
+\frac{4{\pi}g^2{v_g}}{v(\Delta_p+i\Gamma/2)}\Psi^{\dagger}\Psi^2
+\textrm{noise}. \eea
Identifying $\Gamma_{1D}=4\pi{g^2}/v$ and ignoring the loss terms
reproduces the ideal NLSE given in Eq.~(\ref{eq:NLSE}).

We now consider carefully the limits under which
Eq.~(\ref{eq:NLSE}) well approximates the complete dynamics of the
field. First, one requires that the ac Stark shift of $\ket{c}$
due to the nonlinear interaction fits within the frequency range
where EIT is efficient~(\textit{i.e.}, within the transparency
window), which is conservatively satisfied when
$n_{ph}/n_{z}{\ll}|\Delta_p|/|\Gamma-2i\Delta_0|$. In addition,
requiring that higher-order derivatives of the field be negligible
compared to those appearing in Eq.~(\ref{eq:NLSE}) places a
restriction on the maximum wavevector $k_{max}$ of the spin-wave
excitation.  In the TG regime, for example, $k_{max}{\sim}n_{ph}$
and one consequently finds that
$n_{ph}/{n_z}{\ll}\Gamma_{1D}/|2\Delta_0+i\Gamma|$. Finally, as
discussed further in SI, one must also ensure that the loss terms
in Eq.~(\ref{NLS}) do not cause the dissipation of too many
photons, which sets a maximum allowed evolution time $t_{max}$ for
the system.

\subsection{Density-density correlations in one-dimensional system of bosons}
Evaluation of the density-density correlation function in the
ground state is challenging even though the Hamiltonian in
Eq.~(\ref{LL}) is exactly solvable. However, in the regime of
interest where $\gamma$ is large, one can obtain an analytic
expression~\cite{BIK} (we assume that translational invariance is
present and therefore
$g^{(2)}(z_{1},t_{1};z_{2},t_{2})=g^{(2)}(z_{1}-z_{2}=\Delta
z,t_{1}-t_{2}=\Delta t)$),
\bea g^{(2)}(\Delta z,\Delta
t)=1+\frac{K}{4\pi^{2}n_{ph}^2}\int_{|q_{1}|>\pi
n_{ph}}\int_{|q_{2}|<\pi n_{ph}}e^{i
t(q_{1}^{2}-q_{2}^{2})}\cos[(q_{1}-q_{2})z\sqrt{K}][1+\frac{q_{1}-q_{2}}{\pi
\tilde{g}}\int_{-\pi n_{ph}}^{\pi
n_{ph}}(\frac{1}{q_{3}-q_{1}}-\frac{1}{q_{3}-q_{2}})], \eea
where $K=1+4/\gamma$. Note that $g^{(2)}$ decays as $t^{-1}$ for a
range of $t>0$.

For arbitrary interaction strength one can use the exact solution
to numerically evaluate $g^{(2)}$~\cite{BIK,CC,CC2}. On the other
hand this numerical solution~\cite{CC,CC2} as well as a number of
other arguments~\cite{Haldane,Giamarchi} suggests that an
effective description of a one-dimensional gas of bosons, the
Luttinger liquid theory, provides very good long-distance,
long-time behavior. Using this theory, one can demonstrate that
the density-density correlation function in the ground state
exhibits a power-law decay of correlations and $2k_{F}$ Friedel
oscillations, \bea g^{(2)}(\Delta z,\Delta
t)=1+\frac{K}{2\pi^{2}n_{ph}^2}\frac{(\Delta z)^{2}-(v_{g}\Delta
t)^{2}}{[(\Delta z)^{2}+(v_{g}\Delta t)^{2}]^{2}}+\frac{B
\cos(2k_{F}\Delta z)}{n_{ph}^2|\Delta z+iv_{g}\Delta t|^{2K}} \eea
where $B$ is a non-universal constant. The interaction parameter
$K$, or Luttinger parameter~\cite{Haldane,Giamarchi}, can be
numerically extracted from the exact solution. In the limit of
strong interactions, $K=1+4/\gamma$ and tends to 1 in the TG
regime. Here $k_{F}=\pi n_{ph}$ is an emergent {\it Fermi}
momentum.

In the case of sudden switch on of the interaction, one can use
the completeness of the Bethe-Ansatz wave functions basis and
expand the initial state over this basis. Since the matrix
elements of the density operator in the Bethe states are known
\cite{Sl}, computation of the density-density correlation function
starting from an arbitrary initial state can in principle be
performed~\cite{RDG}.

\subsection{Expansion of the pulse}

We consider the expansion of the optical pulse during stages i)
and ii) introduced in the Section ``Preparation and detection of
strongly correlated photon gas''. In regime i), the interaction
energy of particles can be neglected in comparison with the
kinetic energy as long as $\tilde{g}(t)n_{ph}$ is smaller than the
kinetic energy $1/m_{\scriptsize\textrm{eff}}z_0^2$.  The time
$t_1$ at which these energies become comparable satisfies
$t{\propto}-\log (N_{ph}^2\gamma_0)$, and thus if one chooses
parameters such that $N_{ph}^2\gamma_0>1$ free expansion can be
ignored~(for large photon number one can simultaneously satisfy
$\gamma_{0}{\ll}1$ such that the system is initially
non-interacting). Regime ii), which is valid until
$\gamma(t=t_c)=1$, can be analyzed using the usual hydrodynamic
equations (see, \textit{e.g.}, Refs.~\cite{CK,CK2}). From the
equation of motion for the flow velocity, we find $\dot{v}(t) \sim
\frac{n_{ph}}{m_{\tiny\textrm{eff}}z_0} g(t)$, and from the
conservation of particle number, $\dot{n}(t) \sim \frac{n_{ph}}{
z_0} v(t)$. Assuming that the change in density at the center of
the pulse is small, these equations can be integrated to yield a
relative change of density $\Delta
n_{ph}/n_{ph}{\sim}1/\beta^{2}N_{ph}^2$.

\begin{figure*}[p]
\begin{center}
\includegraphics[width=8cm]{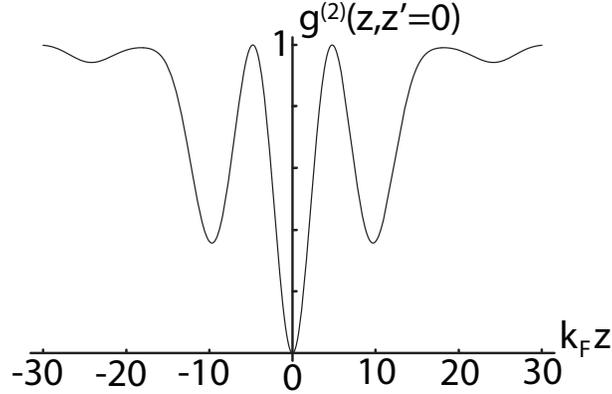}
\end{center}
\caption{\textbf{Density-density correlation function}
$g^{(2)}(z,z'=0)$ for an expanding Tonks-Girardeau gas of photons
with initial density profile $n_{ph}(z)=n_0 (1-z^2/z_0^2)^{1/2}$.
This expansion is equivalent to the problem of a TG gas released
from an initial parabolic confining potential. $z'=0$ denotes the
center of the pulse, and distances are indicated in units of
$k_{F}^{-1}$. The density-density correlation function shown here
is for a system of $N_{ph}=10$ photons, $z_0{\approx}5k_{F}^{-1}$,
and at a time $t=10\omega_{F}^{-1}$ following the initial
release.\label{fig:TG}}
\end{figure*}

Next we consider the expansion in the strongly interacting regime.
Recent work~\cite{muramatsu,MG} demonstrated that spreading of a
pulse of hard core bosons enhances the ``fermionic'' character of
the wavefunction~\cite{muramatsu}. While generally this problem
requires numerical simulations, the solution turns out to be
extremely simple for an initial pulse density of the form
$n_{ph}(z)=n_0 (1-z^2/z_0^2)^{1/2}$, which corresponds to hard
core bosons released from a parabolic potential~\cite{MG}. In this
case the structure of the density correlations is preserved and
there is only a rescaling of length scales~\cite{MG}. An explicit
calculation of $g^{(2)}$ for this system is shown in
Fig.~\ref{fig:TG}. Another signature of fermionization of an
expanding pulse of hard core bosons can be observed in the
momentum distribution function $n(k)$ (density at wavevector $k$),
corresponding to the Fourier transform of the first order
correlation function $g^{(1)}(z,z')= \langle \psi(z) \psi^\dagger
(z') \rangle$. After sufficiently long expansion it approaches a
Fermi distribution~\cite{muramatsu,MG}.  We emphasize that the
qualitative features of our results remain valid for various
shapes of the initial pulse, including Gaussian and parabolic
pulses, although numerical analysis is needed to obtain precise
answers.


\begin{acknowledgements}
We gratefully acknowledge support from the NSF, Harvard-MIT CUA,
DARPA, Air Force, and Packard Foundation.
Correspondence and requests for materials should be addressed to
E.A. Demler.
\end{acknowledgements}

%

\end{document}